\documentclass[sigconf]{acmart}

\usepackage{booktabs} 

\usepackage[utf8]{inputenc}
\usepackage[T1]{fontenc}

\renewcommand\footnotetextcopyrightpermission[1]{} 
\usepackage{listings}
\usepackage{xcolor}

\colorlet{punct}{red!60!black}
\definecolor{background}{HTML}{EEEEEE}
\definecolor{delim}{RGB}{20,105,176}
\colorlet{numb}{magenta!60!black}

\lstdefinelanguage{json}{
    basicstyle=\normalfont\ttfamily\scriptsize,
    numbers=left,
    captionpos=b,
    numberstyle=\scriptsize,
    stepnumber=1,
    numbersep=8pt,
    showstringspaces=false,
    breaklines=true,
    frame=lines,
    backgroundcolor=\color{background},
    literate=
     *{0}{{{\color{numb}0}}}{1}
      {1}{{{\color{numb}1}}}{1}
      {2}{{{\color{numb}2}}}{1}
      {3}{{{\color{numb}3}}}{1}
      {4}{{{\color{numb}4}}}{1}
      {5}{{{\color{numb}5}}}{1}
      {6}{{{\color{numb}6}}}{1}
      {7}{{{\color{numb}7}}}{1}
      {8}{{{\color{numb}8}}}{1}
      {9}{{{\color{numb}9}}}{1}
      {:}{{{\color{punct}{:}}}}{1}
      {,}{{{\color{punct}{,}}}}{1}
      {\{}{{{\color{delim}{\{}}}}{1}
      {\}}{{{\color{delim}{\}}}}}{1}
      {[}{{{\color{delim}{[}}}}{1}
      {]}{{{\color{delim}{]}}}}{1},
}

\newcommand{\mypara}[1]{\vspace{5pt}\noindent{\bf {#1}}}

\begin{document}

\title{DClaims: A Censorship Resistant Web Annotations System using IPFS and Ethereum}

\author{João Santos$^{\dagger}$, Nuno Santos$^{\dagger}$, David Dias$^{\ddagger}$}
\affiliation{%
  \institution{$^{\dagger}$INESC-ID / Instituto Superior Técnico, Universidade de Lisboa\\$^{\ddagger}$Protocol Labs}
}
\email{joao.marques.santos@tecnico.ulisboa.pt, nuno.santos@inesc-id.pt, david@protocol.ai}

\begin{abstract}
The proliferation of unreliable and biased information is a significant problem on the Internet. To assess the credibility of the information retrieved from news websites and other sources, users often resort to social platforms looking for confirmation with trustworthy parties. However, users may be faced with considerable obstacles posed by the platform provider, who can prevent access to certain content. This paper presents DClaims, a system that provides a censorship-resistant distributed service for the exchange of information over the Internet using web annotations. DClaims' fully decentralized architecture relies on Inter-Planetary File System (IPFS) and Ethereum blockchain, both of which offer desirable censorship resistant properties. DClaims is implemented as a web annotations browser extension which allows for the classification of news articles, on news websites. From our evaluation of the system, we conclude that a large scale implementation of the system is practical and economically viable.
\end{abstract}

%
%

\maketitle


\section{Introduction}
\label{chap:intro}

The web plays a fundamental role in the exchange of information in the modern society supported by social media and news websites~\cite{Cushion:2016gv}. Ironically, flooded by a deluge of information and plagued by the propagation of false or biased news, individuals have experienced growing difficulty in accessing reliable sources of information -- to the point of the identification and classification of low-quality information has become one of the most active areas of research~\cite{Buntain:2017by,Kim:2018be,Wu:2018br,Sethi:2017bb}.
As a result, to access information and validate its credibility, people tend to resort to social networks and similar platforms to read posts from reliable sources, exchange commentaries with trustworthy parties, access first-hand content, or cross-check information that appears in news outlets.

However, users' exposure to new messages is largely controlled by platform providers like Facebook, Twitter, or Instagram. Even though the users have the freedom to select who to follow in the network, the platform provider decides on the contents that each user sees based on heuristics and proprietary algorithms. On the other hand, anecdotal evidence suggests that platform providers are prone to pressure by political or economical agents, and sometimes are driven by ideological motivations to hide messages or block certain users. In fact, several instances of censorship treatment of information have recently come to light involving major social network providers, either motivated by bias~\cite{Robertson:2018vj}, or attacks by platform users who flag content as abusive in an attempt to force the platform to delete it~\cite{Tena:UR4pG4AQ}. Ironically, even though the original data may remain accessible to end-users via direct URLs, the constant sharing of new information and the increasing dependency on social networks for content dissemination tends to narrow down users' world view which is molded according to the information displayed to them by the platform and will ultimately impair users' ability to freely and openly access important information. As such, our goal is to study an alternative approach to social networks as a way to offering users unrestricted access to information. 

Over the last few years, blockchain~\cite{Anonymous:JOJGrvgg} has emerged as a powerful technology to avoid the dependency on centralized parties for preserving the integrity of globally shared state. Used for multiple applications~\cite{Dapps:cEzXcv-U}, we hypothesize that blockchain can also serve as a corner stone for overcoming the censorship policies potentially implemented by social networks and other centralized information relay platforms. In particular, we argue that by leveraging a blockchain for building a reliable and open index of messages between producers and consumers of information, then centralized platform providers will not be able to exert their influence in eclipsing undesirable messages from end-users.

Using blockchain for such a purpose, however, poses several technical challenges. First, we face scalability issues since blockchain technology cannot sustain a high transaction rate. Second, the costs tend to be high, because transactions involve a monetary price and an attractive incentive model must be found in order to foster the deployment of such a system. Storage in the blockchain is also a problem, given that the transaction costs tend to increase with the amount of data that gets stored. It is also necessary to deal with potential adversaries aiming at polluting the blockchain with useless messages in an attempt to make the relevant messages more difficult to retrieve by the end-users. The performance is also an important requirement because submitting new messages and retrieving them must not introduce such overheads that will dramatically hinder the user experience. Lastly, the mechanisms for exchanging messages and make them reach the end-users should not depend on existing social networks or require changes to the original news outlet sites while at the same time offering a user-friendly interface.

In this paper we address these challenges by designing, implementing, and evaluating a system named DClaims. DClaims provides a censorship-resistant distributed service for the exchange of information over the Internet using web annotations, which allow us to exchange messages in a user-friendly manner in a fully distributed fashion. Web annotations~\cite{Consortium:PV3E--VD} are defined in a W3C standard, and offer a new way of interacting with information on the web that empowers users to highlight text on the web sites they visit, create sticky notes or comment specific parts of a web page, and share it with friends. In a typical scenario, a user visiting a news webpage article that shows portions of text highlighted by her friends and when she places her mouse over text highlights she sees comments made by the users about that text. Web annotations allow for the creation of a new layer of data, on top of the existing websites, without changing the original resources.

To provide open and censorship-free access to web annotations, DClaims features a fully decentralized architecture based on several components. First, we use the Ethereum~\cite{Buterin:2013ux} blockchain to keep a permanent, canonical record of all the annotations made. These records that are assured to be fresh and ordered. Given that this blockchain is very large and entirely decentralized, it is very difficult to be controlled by any government authority or media outlet. These properties guarantee that every time a user queries the system, he is receiving the latest information, unfiltered and uncensored.
Second, we use Inter-Planetary File System (IPFS)~\cite{Benet:2014vw}, a distributed file system where data is stored and scattered all across the web. 
IPFS offers strong integrity assurances of the files it stores. That happens because the link to an IPFS file depends on the content of the file rather than the file's location. To combine the feature set of web annotations, with the integrity and censorship resistance assurances of IPFS and the ordered registry and freshness of Ethereum, our system stores web annotations on IPFS and records the IPFS links of these files on Ethereum. Most importantly, DClaims includes additional components, named {\em publishers}, which reduce the costs of Ethereum transactions and improve the system scalability. Publishers incorporate mechanisms to prevent publisher misbehavior and to protect against spurious content.

We have implemented and evaluated DClaims. We found that the system performs well, in particular that the web browsing experience of end-users is not significantly affected. We also analyzed the costs of a full-scale deployment of DClaims using the activity level of Facebook's news pages as an estimate for expected demand which suggests that our system is economically viable.


\section{Building Blocks}
\label{sec:motivation}

The goal of our work is to build a system that allows for uncensored access to web annotations on the Internet by eliminating central points of control where a powerful actor can exert pressure to fabricate, modify, or suppress exchanged messages. Examples of such actors include governments or powerful media organizations.
In addition, our system must provide: authenticity and integrity assurances regarding the published web annotations, data permanence and portability with respect to immutability of links to web annotations, financial cost efficiency, and scalability.

We present the three main building blocks that we leverage to build our system: Web Annotation Data Model, Ethereum, and IPFS.

\subsection{Ethereum}
\label{rel:bitcoin}
\label{rel:blockchain}

The Ethereum~\cite{Buterin:2013ux,Wood:2014ur} blockchain is the core building block of DClaims. It consists of a blockchain, which is a distributed system that runs on a peer-to-peer network with the goal of maintaining common state in a trustless manner, that is, without trusting in any single node to act correctly. Agreement on the state is achieved through a consensus protocol, based on proof-of-work.

Ethereum's purpose is to run a global virtual machine on the blockchain that anyone can use by paying a small fee. 
This virtual machine runs programs, called {\em smart-contracts} written in Solidity and compiled into EVM Bytecode. All the nodes in the Ethereum network run the same operations. Since the virtual machine is deterministic (for a given input, the output is always the same), all nodes will reach the same state, which results in the network achieving consensus. Smart-contracts are Turing-complete which provides flexibility and allows for the creation of arbitrary programs.

Ethereum exhibits strong censorship resistance features: 
Ethereum nodes are geographically spread and controlled by different parties. However, important limitations need to be overcome. 
The first challenge is minimizing the data stored on the blockchain, for it is extremely expensive. The second challenge is minimizing the number of transactions so that Ethereum's 20 transaction per second limitation does not turn into a bottleneck in the system.

\subsection{IPFS}
\label{rel:ipfs}

As described, storing data on a blockchain is expensive. IPFS offers a good solution to that problem. 
The Interplanetary File-System (IPFS)~\cite{Benet:2014vw} is a decentralized, peer-to-peer file system. The IPFS network is made of IPFS nodes. An IPFS node is a client running the IPFS daemon, which stores files in its local IPFS repository (a datastore used by the IPFS daemon) and makes those files available to the rest of the network. 

Data is addressed by IPFS links, which are hashes of the content they address. Routing is done through a distributed hash table. Transport is done by establishing direct, peer-to-peer, connections between nodes.
IPFS offers desirable features for censorship resistance. First, it is logically decentralized, meaning it can work in local area networks, disconnected from the Internet. Second, it does not rely on DNS or Certificate Authorities. Third, even if only one node in the network has the requested file, all the other nodes can access it. Finally, all files are cryptographically verified.

\subsection{Web Annotation Data Model}

To allow for the decentralized exchanging of messages between users without depending on social network platforms or requiring changes to news web sites, we leverage the web annotation standard specified by the W3C~\cite{Consortium:PV3E--VD}. It specifies the creation of a new layer of data, on top of the existing websites, without changing the original resources. This layer can be used to provide context, clarification, additional information about the resource a person is viewing.

Several types of representations for web annotations are supported, such as text or media (sound, images or videos). The target can be any HTML element in the webpage. 
In our work, we adopt the W3C web annotation specification in order to preserve interoperability between our system and other web annotation services.


\begin{figure}[t]
  \centering
  \includegraphics[width=0.9\columnwidth]{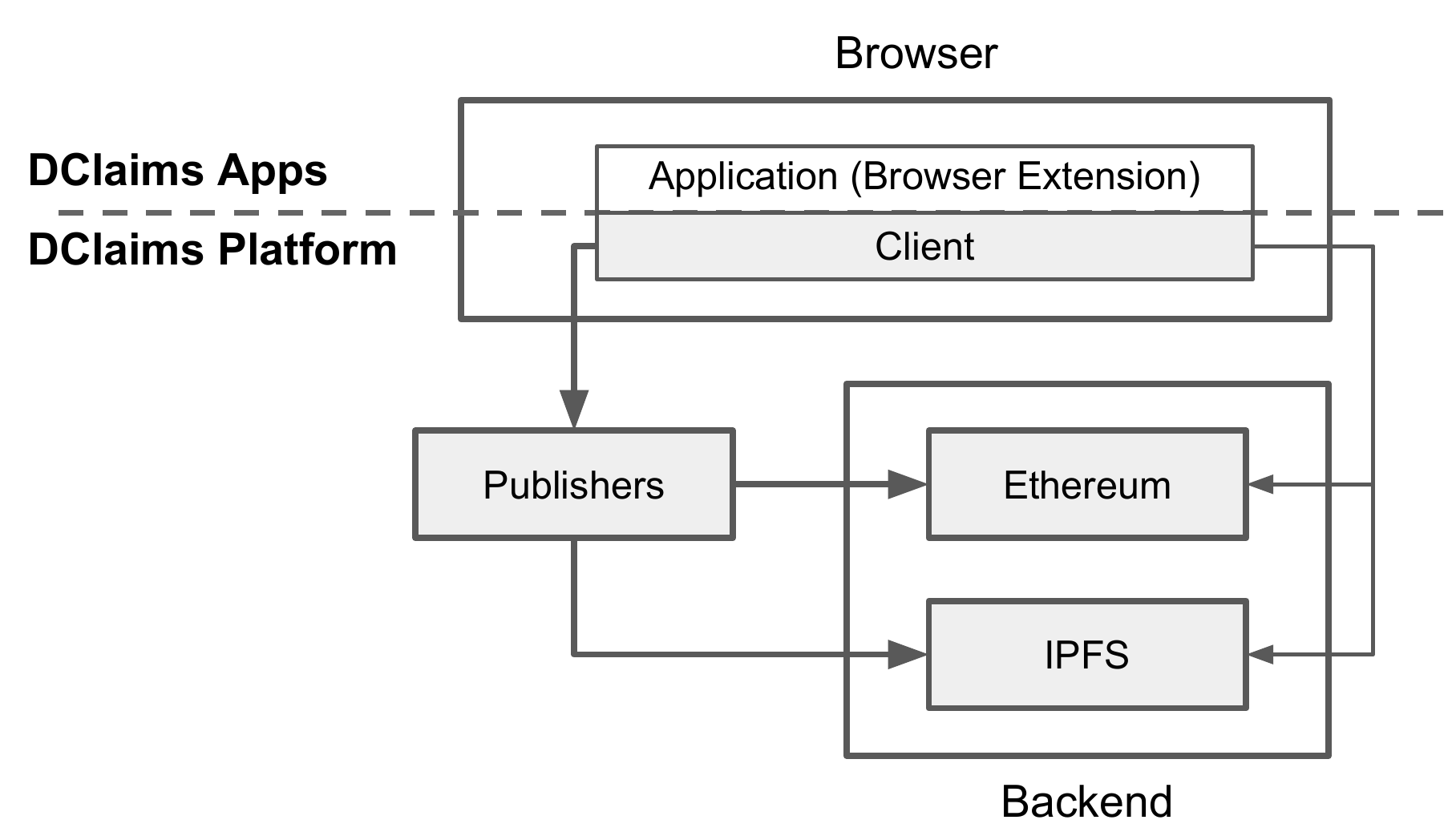}
  \caption{DClaims architecture}
 \vspace{-0.4cm}
  \label{fig:archoverview}
\end{figure}

\section{Design}
\label{chap:architecture}

We introduce DClaims. We start by presenting its architecture, and then describe the most relevant aspects of its design in turn.

\subsection{Architecture}

Figure~\ref{fig:archoverview} represents the architecture of our system. Essentially, DClaims consists of four main components: client, publisher, IPFS, and Ethereum. Together, these components constitute the DClaims platform on top of which applications can be built. By leveraging the underlying platform services, applications allow end-users to create, store and share web annotations in a decentralized and censorship-resistant way.

More specifically, a DClaims application consists of a browser extension that runs on end-users' browsers. Depending on its specific functionality, an application may dynamically instrument web pages so as to extend their respective UIs with web annotation widgets, or provide such features through an API that web pages can explicitly use. The application code implements specific logic for generating new annotations, submitting them to the DClaims platform, retrieving, and displaying them to the end-user. Multiple applications are allowed to use the system.

To interact with the DClaims platform, an application must be linked with a library named {\em client}. The client provides a simple API and implements common logic for the submission of requests to the backend for storing annotations and make them accessible to DClaims' users. The backend consists of IPFS and Ethereum blockchain and is responsible for providing a censorship-resistant storage and access medium for the system's state, namely annotations and metadata. According to the instructions of the application, the client can submit annotations to backend either directly by communicating with IPFS and Ethereum, or indirectly by relying on a special DClaims proxy named {\em publisher}. There can be multiple publishers. Publishers consist of servers maintained by third-parties and are responsible for issuing annotations on clients' behalf for cost efficiency and scalability reasons, which will be clarified later in the paper. The client also implements common logic for user authentication and cryptographic validation of annotations.

\begin{figure*}[t]
  \centering
  \includegraphics[width=1.6\columnwidth]{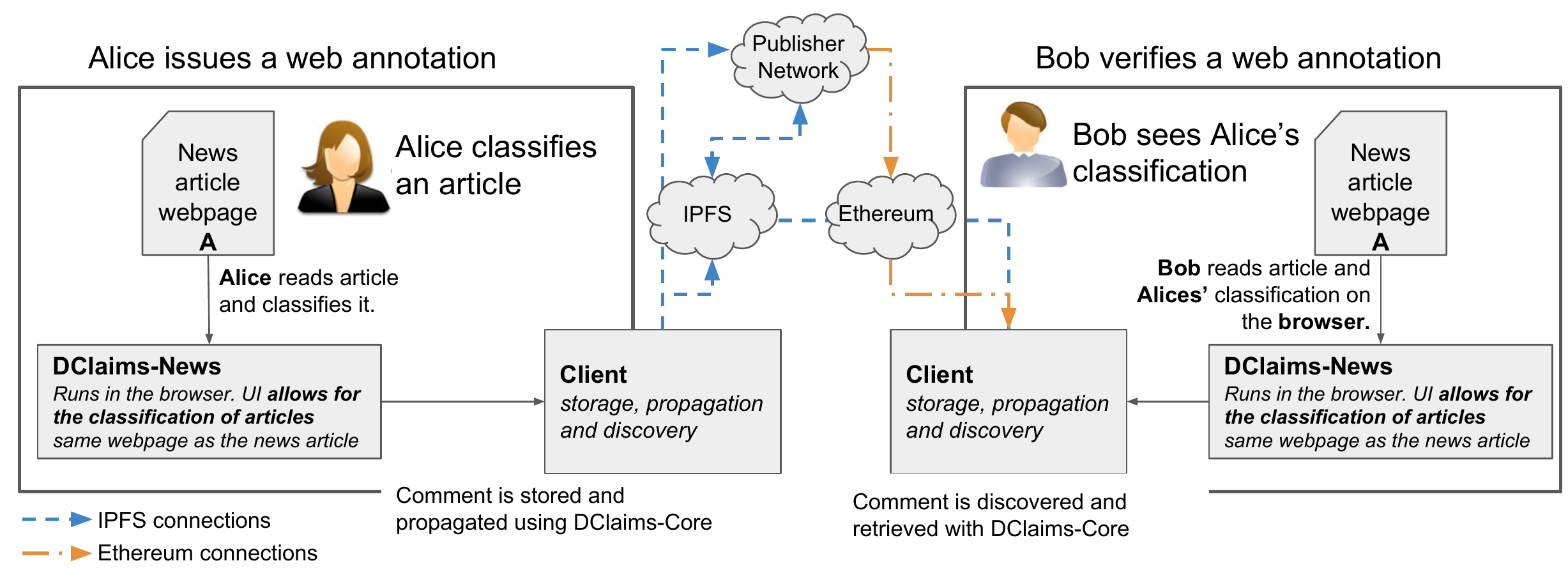}
  \caption{DClaims-News basic workflow: Alice writes a web annotation on a news article which is then read by Bob.}
  \label{fig:arch}
\end{figure*}

\subsection{Assumptions and Threat Model}

The security goals of our system are threefold: protect the integrity of exchanged web annotations, guarantee unrestricted access to them by end-users, and provide strong authenticity assurances.

To this end, we make several assumptions. First, we trust the client endpoints. We assume that the application, client library, browser, and OS are correct, in particular with respect to the implementation of DClaims protocols and security of cryptographic keys and operations. As part of the trusted computing base (TCB) we also include Ethereum, which requires that at least 50\% of the hashing power to operate correctly. We depend on Ethereum for guaranteeing the integrity of the execution of DClaims smart-contracts and preserving the freshness of the smart-contracts' internal state, e.g., by preventing rollback attacks. With regards to IPFS, DClaims uses it for state replication and only requires a single data copy to be preserved on any given IPFS node to ensure data persistence and accessibility. The integrity of data objects stored on IPFS can be verified by the clients. Lastly, we assume that the publisher may deviate from the correct behavior. If such happens, the overall censorship-resistance guarantees offered by the system must not be affected but temporarily. Thus, we exclude the publisher from the TCB.

We model our adversary as follows. The adversary may try to intercept and / or tamper with any communications taken place between parties of the DClaims ecosystem. The adversary may attempt to impersonate a legitimate publisher or deploy a malicious publisher. Such a publisher could launch several attacks, such as altering or dropping annotation requests submitted by clients, which would result in either modifying the content of legitimate annotations or preventing them from being persistently stored on the backend. Note that we are not contemplating an attacker model which actively engages in censorship techniques at the transport level (such as IP blocking, packet dropping or content inspection).

\subsection{System Operation}
\label{sec:sysop}
\label{dclaimsnews}

We start the description of DClaims by presenting its basic operation from the end-user's point of view, which is centered around the life cycle of web annotations. Throughout its life cycle, a web annotation is handled by four entities: \textit{creator}, \textit{issuer}, \textit{verifier} and \textit{viewer}. The creator is responsible for the creation of a new web annotation. The issuer submits the annotation to the backend. A verifier retrieves the annotation from the backend and verifies its authenticity. Finally, a viewer displays the annotation on a web page. For each annotation, there can be multiple verifiers and viewers.

Figure~\ref{fig:arch} illustrates this process for a simple application, named DClaims-News, whose goal is to classify online news articles. This example involves two users Alice and Bob, both of them have installed DClaims-News on their browsers. First, Alice visits a news website and reads an article. DClaims-News instruments the article's web page so as to allow Alice to classify the information therein contained with one of two possible tags: true, or false. Upon receiving the input, DClaims-News (the creator) creates a new annotation based on Alice's choice, and forwards it to a publisher (the issuer), which will submit the annotation to the backend. Later, Bob visits the same article news article website. Upon opening the article's website and Bob visits the respective new's web page, DClaims-News instruments that page on the fly with a button that allows Bob to read the classifications attached to that article, which in the background leverages the local client (the verifier) to retrieve the article's annotations and verify them. When Bob presses the button, DClaims-News (the viewer) displays the existing classifications on a new pop-up window allowing Bob to browse them through and checking who has authored each classification. In this case, he could see Alice's classification of the article. Next, we use this example to present the design details of DClaims.

\subsection{Data Structures}
\label{dclaims-core:main}

DClaims uses three main data structures to manage the system's state. One of these data structures is called \textit{claim} and it is used primarily to encapsulate web annotations provided by applications along with additional metadata, e.g., digital signatures and user identification. To foster interoperability
with other decentralized applications, claims are stored inside Verifiable Claims~\cite{vclaims}, a data specification standard defined by W3C which allows for expressing rich sets of signed statements. DClaims adopts this data model for the representation of claims, which in turn have four different formats for four different purposes: enclosing web annotations, batching publisher requests, revoking claims, and filing publisher complaints. The web annotation claim format can be further customized according to the application's needs, e.g., to represent simple {\em true} / {\em false} statements, or arbitrarily complex ones (e.g., structured records, text, images, etc.). 

\begin{figure*}[t]
  \centering
  \includegraphics[width=1.5\columnwidth]{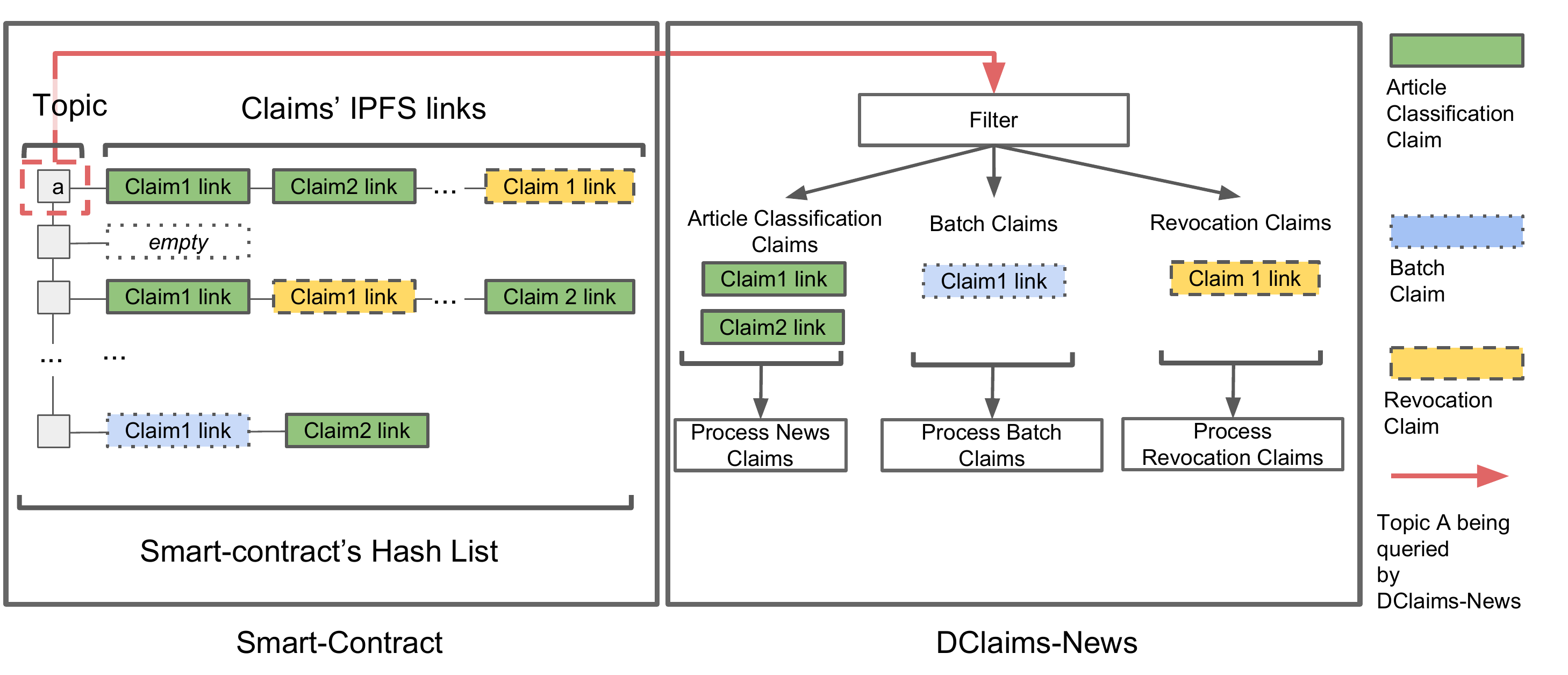}
  \caption{Representation of smart-contract and DClaims-News data structures: article A is queried (topic A in the hash list).}
  \label{fig:hashlist}
\end{figure*}

The second important data structure is an Ethereum smart-contract, whose function is to keep track of the claims issued. Essentially, whenever claims are stored on IPFS, the links (i.e., self-describing content hashes, with hashing function and length information embedded) to these claims are put inside the smart-contract. Figure~\ref{fig:hashlist} represents the smart-contract's data structure that maintains the claims' IPFS links for the DClaims-News application. The smart-contract holds a hash list where the key is named {\em topic}. The list contains the IPFS links, issuer addresses, and time stamps of all the claims about that topic. 
The topic can be represented by an URL, e.g., web annotations for \texttt{https://www.acme.com/index.html} fall under the same topic. In this example, the represented annotations correspond to classifications of the article as being \textit{true} or \textit{false}. 

The third relevant data structure is a second Ethereum smart-contract, whose function is to maintain a directory of all publishers in the system and to keep track of the complaint claims that might have been issued against publishers. Section~\ref{arch:bad-publishers} elaborates further on this topic. To distinguish between both DClaims smart-contracts, we refer to the former as {\em web annotation smart-contract} and to the latter as {\em publisher registry smart-contract}. Since the web annotation smart-contract plays a more prominent role in the system, in the rest of the paper we refer to it simply as ``the'' DClaims' Ethereum smart-contract unless stated otherwise.

\subsection{Claim Management Operations}
\label{arch:apps-verification}
\label{dclaims-news:verification}

We now revisit the web annotation operations of Section~\ref{sec:sysop} and provide additional details on how they work. We omit further details about the visualization operation because it is 
application-dependent, and introduce a new operation for claim revocation.

\mypara{Claim creation:} For claim creation, we only need to mention that the creator adds a signature of the claim's content. This signature allows the claim's validators to check its authenticity and integrity.

\mypara{Claim issuance:} To issue a claim to the backend, the issuer performs three steps: first, it signs the claim with its private key, then it must store the claim on IPFS so as to obtain the respective IPFS link to the claim, and then insert the link into the hashlist of the smart-contract. When a new claim is issued, the smart-contract uses the issuer's Ethereum wallet address as identity proof, meaning that that wallet address is automatically associated with that claim. Depending on who issued the claim, the wallet address associated with the claim may belong to the publisher or to the user who has created the claim. The smart contract contains a record of the issuer ID which serves to identify the entity who paid for the transaction and issued the claim. On the other hand the ID on the claim itself identifies the user who has provided the enclosed web annotation.

\mypara{Claim revocation:} DClaims allows for revoking previously issued claims. Revoking a claim consists of issuing a special {\em revocation claim}, which is tagged with revocation type and has a payload that includes the UID of the claim to be revoked. This claim is issued just like any other web annotation claim, either by a client or by a publisher; it is stored on IPFS and kept track of on the Ethereum smart-contract (see Figure~\ref{fig:hashlist}). To prevent an attacker from revoking claims arbitrarily, a revocation claim must have been created (and signed) by the creator of the claim to be revoked. To enforce this condition, rather than performing this check inside the smart-contract (which would be very costly and non-scalable), we allow anyone to submit arbitrary revocation claims into the system and let the test be performed at the verification stage by the clients. To avoid attackers from flooding the system with bogus revocation claims, DClaims implements a defense mechanism described in Section~\ref{sec:spam}. Lastly, note that, since IPFS does not enforce access control to stored objects, once a given claim has been uploaded onto IPFS, it can be accessed and replicated by anyone in the network. For this reason, DClaims does not provide a mechanism for deleting claims, only for revoking. Nevertheless, if end-users intend to share private claims, this can be achieved by encrypting the claims' payload and sharing the decryption keys with trusted users.

\mypara{Claim verification:} This operation is invoked by the application before displaying the annotations to the user and requires several steps.
First, the client needs to retrieve the claims of a particular topic. To this end, it queries the smart-contract to get the IPFS links of the claims associated with a topic indicated by the application and then fetches the respective files from IPFS. In this process, after acquiring the claims' IPFS links from the smart-contract, IPFS performs a first level of verification to check that the link (hash) of the claim matches its content. 
Upon receiving the claims, a second step of verification is to filter the claims. Filtering allows end-users to select whose claims they want to see, so all the claims from issuers who are not white-listed by the user are discarded. Filtering also helps evict spurious claims (see Section~\ref{sec:spam}). The third step of  verification is to check the claim's signature. The client checks that the public key used to sign the claim is the same as the one in the Issuer ID field. The last verification step consists of ensuring that the claim has not been revoked. If all these checkpoints are passed, the claim is considered valid and passed on to the application. The order through which the verification is made was deliberately chosen with performance in mind.

\subsection{The Publishers Network}

Issuing a claim by uploading its content to IPFS and respectively updating the smart-contract on Ethereum can be done directly by the client. However, this approach can have several shortcomings. First, the claim issuing costs are very high, because the claim creator needs to pay the price of a single Ethereum transaction in order to update the smart-contract's state with the claim's IPFS link, price which is expensive (about 0.25USD per claim).

Second, this method may hinder the scalability of DClaims. The reason is that, as the rate of created claims increases, the Ethereum network will eventually reach a limit on the number of transactions that can be processed. The current limit is around 20 transactions per second, creating a potential bottleneck in the DClaims system. Moreover, by reaching this limit, the gas price spikes. This effect occurs because since transaction prices tend to escalate as the transaction rate approaches the limit.

A third issue is a limited claim availability. After issuing a claim, only a copy of that claim exists on the creator's computer, which acts as an IPFS node. However, IPFS does not automatically replicate the data across the network; it is only when IPFS nodes request the content, that new copies are generated and cached in the IPFS network. As a result, the claim creator would need to be permanently connected to the Internet until a sufficient number of claim replicas had been disseminated across the IPFS network, otherwise end-users may not be able to retrieve the claim.

To overcome these limitations, DClaims provides an alternative, and preferable, way for indirect claim issuance based on a network of dedicated nodes called publishers. Publishers act as proxies between applications and the DClaims backend systems, i.e., IPFS and Ethereum. To address the aforementioned shortcomings, a publisher can proxy two kinds of requests by applications: requests for claim issuance, and requests for claim retrieval. First, when receiving issuing requests from multiple clients, a publisher can batch them into a single Ethereum transaction, which not only reduces the price per claim, but also reduces the overall number of transaction requests to Ethereum thereby improving scalability. Second, a publisher can also act as an IPFS node serving as a storage provider for claim copies serving content on behalf of claim creators.

\subsection{Incentives for Publishers}
\label{arch:publishers-economic}

We envision that publishers will be deployed and maintained by third-party {\em publisher providers}, who must support the maintenance costs of the system, including the costs of Ethereum transactions responsible for serving claim issuing requests and the claim storage costs. Thus, publisher providers require an attractive incentive model and a source of revenue that can cover these expenses. Different publisher providers can adopt different financial models. We discuss two possible models for supporting the publisher network:

\mypara{Donations:} A viable option is to run a community-funded campaign to subsidize the publisher network. 
In Section \ref{eval:cost-publisher}, we estimate that the cost of running a full-scale deployment of DClaims would be 5 to 6 times lower than the one of Wikipedia~\cite{Wikipedia:YoXphYji}, which adopts a donations-based financial model.

\mypara{Pay-as-you-go:}
An alternative model is to have users pay for issuing their web annotations. For instance, supposing that a publisher batches 100 claims from 100 users into a single Ethereum transaction priced 0.25USD, each user would pay 0.0025USD, which seems to be a reasonable price to pay for a censorship-resistant web annotations service.
Publishers can charge regular payments to users (PayPal, credit card), or employ more sophisticated methods, such as requiring users to mine some cryptocurrency on their website. 

\subsection{Operations Involving Publishers} 
\label{arch:publishers-batch}

We imagine the publisher network to be made up of a reasonable number of publishers (between 10 and 100). Different publishers may charge different pricing schemes for claim issuance and storage services. End-users are free to select preferred publisher providers. Next, we present the most relevant operations involving publishers, namely those involved in issuing and storing claims and in managing membership of the publisher network.

When a publisher is used as proxy, the claim issuing process changes in two main ways. First, rather than interacting directly with IPFS and Ethereum, a client sends a claim issuance request to the publisher. The publisher batches claims together and issues them all in the same Ethereum transaction. Publishers can set a threshold on the number of claims to include in a given batch. 
Second, in order to improve the claim availability in the IPFS network, the publisher stores a copy of the claim locally, and propagates similar requests to other publishers according to some pre-defined replication policy (e.g., three randomly distributed copies). Each publisher effectively runs an IPFS node which means that it can serve future claim requests from its local store. The publisher and the claim creator agree upon the storage policy that must be enforced, e.g., preserve a local copy for at least 1 month. Note that the claim creator will always keep a local copy of the claim.

With regard to operations involving publisher discovery and membership management (i.e., add and remove publishers), a simple approach would be to run a public directory service on a web site keeping the IPs and certificates of all publishers of the network. 
However, to prevent depending on a centralized entity which can engage in censorship practices (e.g., by controlling publisher providers' ingress), DClaims keeps the publisher directory on an independent Ethereum smart-contract (the publisher registry smart-contract) where publishers must register. When starting its activity, a publisher makes a transaction to that contract, adding its IP address (or DNS name) and public key certificate to a list. Users and other publishers can find the newly added publisher by querying this smart-contract for the list of active publishers.

\subsection{Handling Publisher Misbehavior} 
\label{arch:bad-publishers}

As explained above, publishers bring significant benefits for DClaims in terms of cost efficiency, system scalability, and data availability. 
However, DClaims does not depend on publishers' correctness for preserving the censorship-resistance properties of the system. These properties rest upon Ethereum and IPFS only. 
As a result, we assume that publishers can misbehave.

\begin{figure}[t]
  \centering
  \includegraphics[width=1\columnwidth]{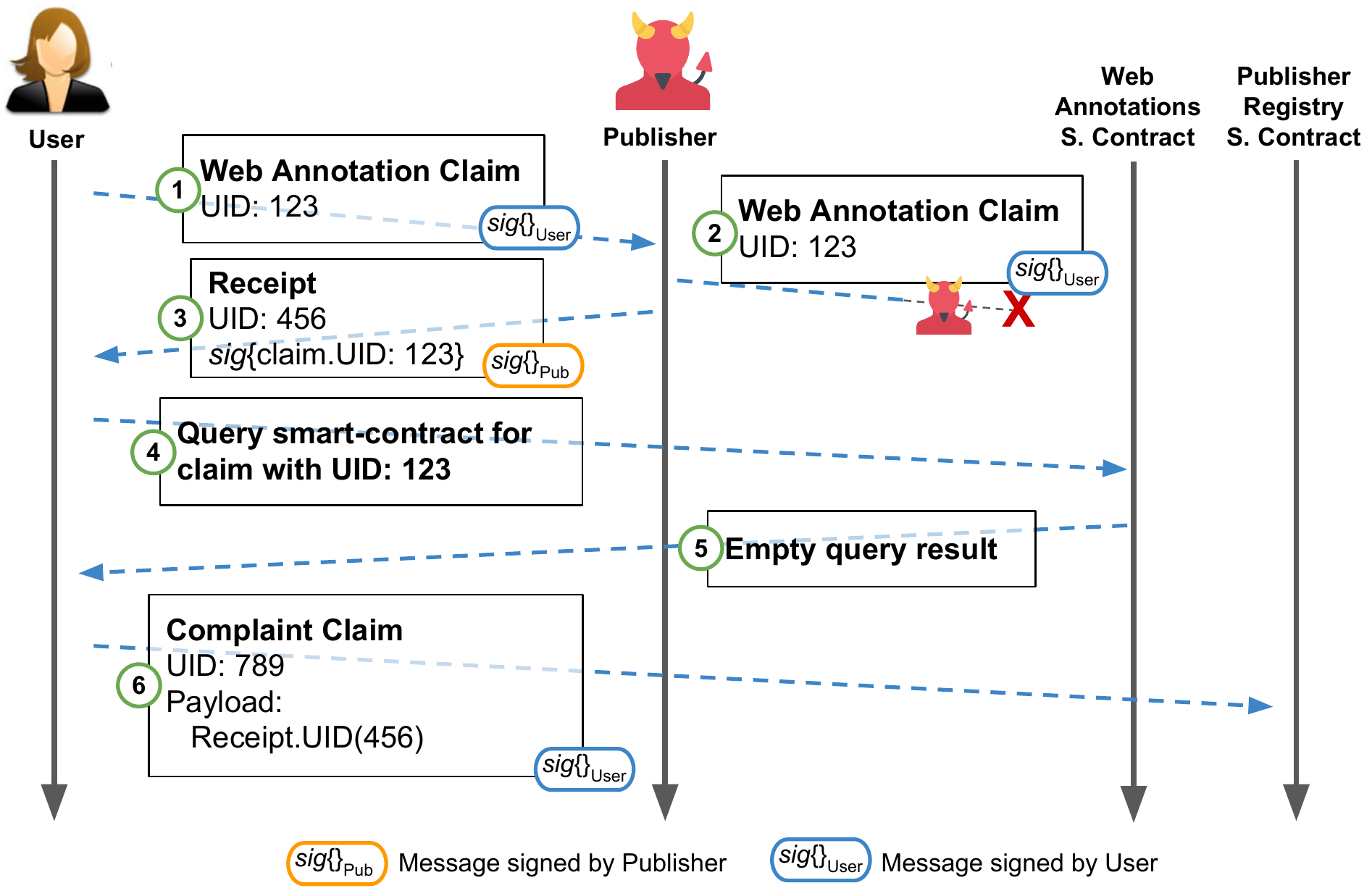}
  \caption{Typical publisher complaint flow.}
 \vspace{-0.4cm}
  \label{fig:dclaims-architecture}
\end{figure}

An adversarial publisher can try to subvert the system in three specific ways: by silently refusing to update the Ethereum smart-contract with the claim's links received from clients for issuance (request corruption), by tampering with or modifying the client's claim issuance request, e.g., registering the IPFS link under a different topic 
(request corruption), and by not storing a copy of the claim submitted for issuance by a client and / or refuse to propagate it further in the publisher network (replica drop).

Although such attacks may cause some disruption in the system, they can easily be detected and recovered from. Request drops and request corruption attacks can be detected by the original claim creator by querying the smart-contract and validate the presence of the correct claim reference. An adversarial publisher could launch such attacks in order to save costs with Ethereum transactions (request drop attack), or to prevent the dissemination of claims due to their content or authorship (which applies to both attack types). A replica drop attack can also be detected by retrieving the claim while keeping the claim creator offline. If the claim has not been served, it means the publisher has not cached a copy of the claim. To recover from these attacks the claim creator may select another publisher to re-issue the claim, or to cache and replicate the claim.

To discourage publisher misbehavior, DClaims incorporates a reputation-based defense mechanism. This mechanism allow users to identify and denounce misbehaved publishers while allowing anyone to attest to the veracity of such reports. The mechanism is based on the generation of \textit{receipts} and {\em complaints} (see Figure~\ref{fig:dclaims-architecture}). When an application requests the issuance of a new claim to a publisher, the publisher returns a receipt. That receipt is a cryptographic proof of the publisher's acknowledgement of the request and promise to fulfill it. If a publisher fails to publish the claim, or denies access to it, the user can file a complaint, and warn other users to stop using that publisher. This information is maintained in the publisher registry smart-contractm, which maintains publishers' identities. Users can validate the complaints in the smart-contract, and disregard false complaints.

\subsection{Protection against Spurious Claims}
 \label{sec:spam}


Since the DClaims platform is public, the system is prone to the potential publication of spurious claims. One possible concern would be that numerous such claims could be generated and pollute the system in such a way as to make it difficult to identify legitimate claims rendering the system useless. We present three DClaims mechanisms that limit the extent of such attacks:

\mypara{1. Paid claim issuance:} Without using the publisher network, an adversary willing to upload spurious claims to the system, needs to pay for Ethereum transaction by itself. The Ethereum transaction price can cost between 0.5 and 1.5 USD, which makes it cost prohibitive to issue a large number of claims for a DOS attack.

\mypara{2. Client authentication:} This mechanism can be implemented by the publisher. A user uses a publisher's service by interacting with its API. A publisher can choose to authenticate and authorize a user in order to allow API usage for issuing claims. If a user issues to many claims and starts abusing the service, a publisher can de-activate that user's account. 

\mypara{3. Client-side whitelisting.}
On the client side, DClaims users are only exposed to claims generated by users that belong in their whitelist. 
To ensure that end-users are not exposed to an excessive amount of claims, users are required to whitelist the issuers whose claims they can see. This way if a user goes rogue and posts spurious claims, other users need only to remove him from their list.


\section{Implementation}
\label{chap:implement}

This section presents the implementation of the developed software stack comprising the base DClaims platform (Section~\ref{sec:implplat}) and the DClaims-News application (Section~\ref{sec:implnews}).

\subsection{DClaims Platform}
\label{sec:implplat}

As part of the DClaims platform, we implemented the client library, the publisher software, and Ethereum smart-contracts. To work with IPFS, we used Go-IPFS~\cite{goipfs}, which is the software that client and publishers use to connect to the IPFS network and to run a local IPFS node. In the client, we also use JS-IPFS-API~\cite{ipfsapi}, which is an IPFS client library written in Javascript and it allows the browser extension to communicate with the Go-IPFS node running locally. (Within a few months from the time of this writing, we will be able to use a Javascript IPFS implementation, which will remove the need for the user to install and run Go-IPFS; the IPFS node will be launched by and run inside the browser.) Both the DClaims' client and the publisher were written in Javascript. The publisher code runs on a Node.js web server.

We wrote the Ethereum smart-contracts in Solidity, which is similar to Javascript in syntax, but typed. To deploy smart-contracts on the Ethereum network, it is necessary to interact with an Ethereum node. On the publisher, we run an Ethereum node in the Node.js environment using the Go-Ethereum~\cite{goeth} (Geth) client. For the DClaims client we used Metamask~\cite{metamask}, which is a browser extension that functions as an Ethereum node proxy. Contrary to Geth, Metamask is not an Ethereum node in that it does not process transactions. It is instead a gateway to Ethereum nodes being run by a third party (Infura~\cite{infura}, which runs public Ethereum nodes). Node.js and Javascript browser applications can connect and interact with an Ethereum node via the Web3 library~\cite{web3}. In Node.js, Web3 connects to Geth and in the browser to Metamask. Browser clients can also choose to run their own Ethereum nodes and not rely on Infura's. To do that, they simply need to configure their Geth clients on Metamask.
 
\subsection{DClaims-News Application}
\label{sec:implnews}

As a proof of concept, we built the web annotation application for news websites introduced in Section~\ref{dclaimsnews}, which allows users to classify, and view classifications, on news articles. This application, named DClaims-News, consists of a browser extension for Chrome and was written in Javascript. It implements a visual overlay that is placed on top of the news websites, which allows users to interact with the application. To draw the visual elements (e.g., buttons to interact with the application) we injected several Javascript files (including the Bootstrap\cite{Bootstrap:9RW7LD06} and jQuery\cite{jQuery:Z1YJbxUl} libraries) via a Chrome browser extension\cite{Google:G3Yt1Wt8} to change the HTML of the web pages.

The interaction of DClaims-News with the DClaims platform is mediated by the DClaims client library. Claim issuance can be performed directly by the application or mediated by a publisher. When issuing a claim, even if using publishers, the users must sign in using its Ethereum Wallet Address. To sign and verify signatures we used the methods made available in the browser environment by Metamask. When a user creates a claim (by classifying an article), a pop-up box from Metamask appears, showing the piece of data the user is about to sign.


\section{Performance Evaluation}
\label{chap:evaluation}

We assess to what extent DClaims can serve as a viable alternative to current web commentary platforms, such as social networks. To that end, we are primarily interested in evaluating the performance of the DClaims-News browser extension (Section~\ref{sec:perffrontend}) and DClaims backend components, namely Ethereum and IPFS (Section~\ref{sec:perfbackend}).

\subsection{Methodology}
\label{eval:ux}

To evaluate the performance of DClaims-News, we have injected Javascript to measure the loading time of the webpage. The measured values correspond to the elapsed time between two events provided by the browser: \verb|requestStart| and \verb|loadEventEnd|. For our tests, each webpage was loaded thirty times on each condition (with and without DClaims running). Our experiments were conducted on a computer equipped with 2.4GHz CPU and 8GB memory connected to a 20Mb network. The web extension was connected to the IPFS network through a daemon running locally on the computer and connected to an Ethereum testnet via Metamask, a web extension that acts as an Ethereum proxy.

To measure the performance of IPFS and Ethereum, we emulated a regular usage by ordinary users by running a DClaims client wrapper on sixty nodes on Amazon Web Services. Each node had 4 GB of memory and had IPFS and Go-Ethereum running on Docker containers. This experiment started by having each node randomly issuing five claims, each about a randomly selected article (from a pre-selected list of thirty, which corresponded to the articles on SkyNews' webpage on January 28th 2018). To avoid the extra burden of configuring a different account on each node, the claims were issued 
using the same Ethereum address. After the claims have been issued, each node started fetching claims, randomly selecting an article 
and retrieving all the claims for that article. Each node selected a new article every 10 seconds. We ran this experiment for twenty minutes, which resulted in each node querying 120 articles. The overlap (querying 120 articles from a list of 60) was intentional to increase the odds of every article being queried at least once.

\subsection{Performance of the Web Client}
\label{sec:perffrontend}

In this section, we report the performance results of our DClaim-News application. To provide an idea of the impact of DClaims to end-users' experience, we adapted our web extension to support three websites: SkyNews~\cite{skynews}, New York Times (NYT)~\cite{nyt}, and Instituto Superior T\'{e}cnico (IST)~\cite{ist}, which is the news front page of a university website.

\begin{figure}[t]
  \centering
  \includegraphics[width=0.75\columnwidth]{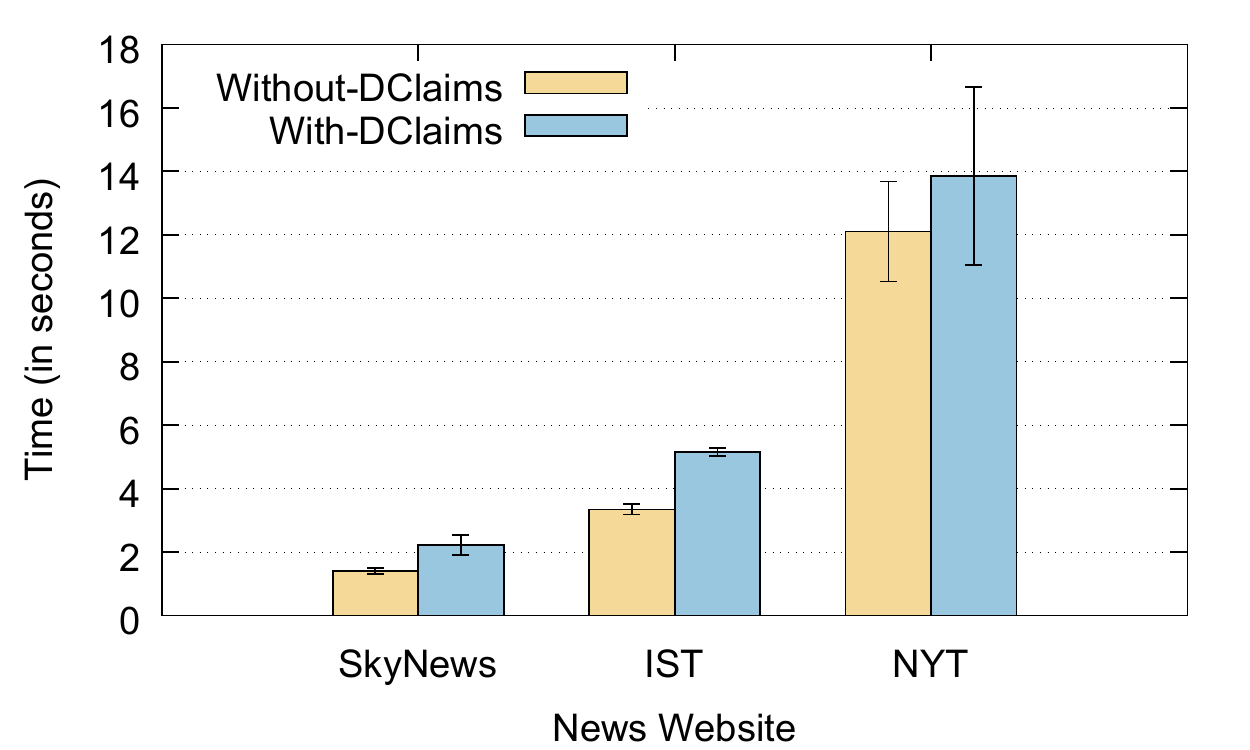}
  \vspace{-10pt}
  \caption{Full website loading times.}
    \vspace{-0.3cm}
  \label{fig:evalwebpage}
\end{figure}
  
Figure~\ref{fig:evalwebpage} shows the time that the three news websites take to load completely, with and without DClaims-News enabled. The overhead DClaims introduced varies between 0.4 seconds and 2 seconds and corresponds to the time that the web extension needs to connect to an IPFS node, to an Ethereum node and then, for each article, it needs to generate the news article ID (which is the SHA-3 hash of the referenced news article's URL). Note, however, that this is not the latency perceived by the user, since this operation takes place in background while the web page is being rendered to the user. From our interaction with these sites using DClaims-News, we did not perceive a dramatic degradation of the user experience, as the original elements of the website (news titles, images, among others) appear just as fast as they did before, only the elements introduced by DClaims (view claims button, claims counter) take longer to appear.

\begin{figure}[t]
  \centering
  \includegraphics[width=0.75\columnwidth]{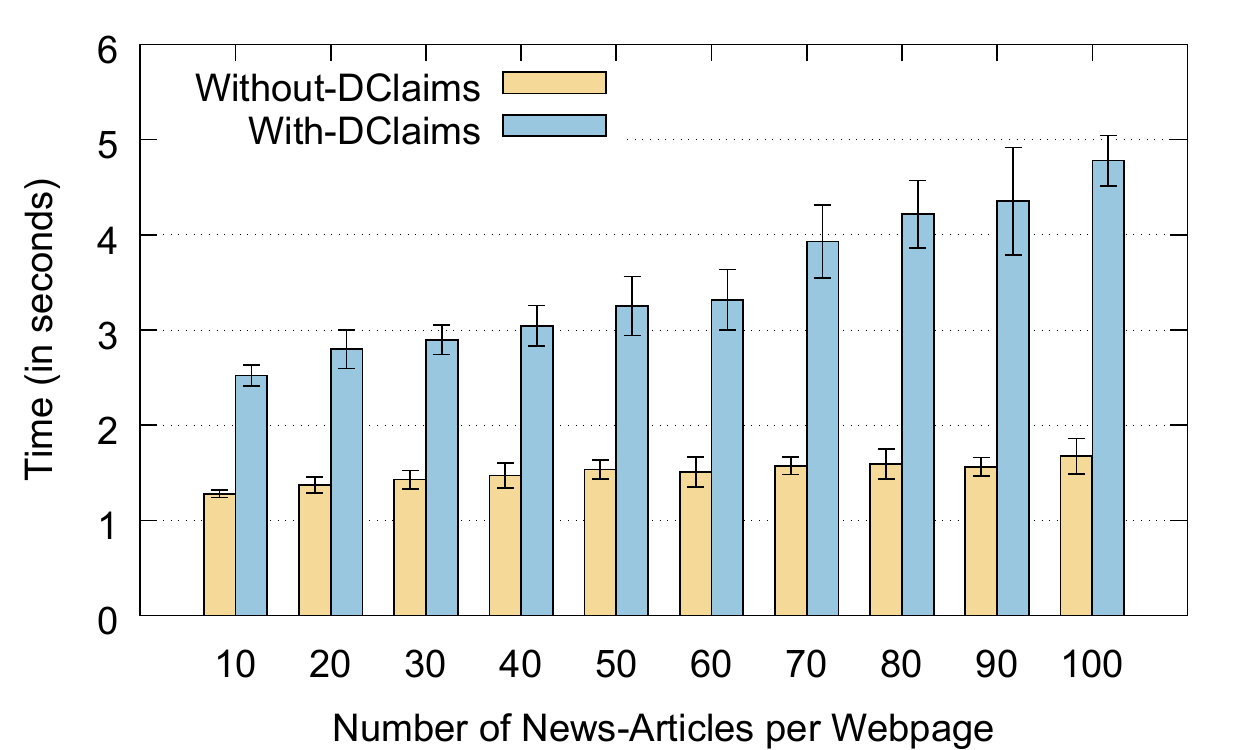}
  \vspace{-10pt}
  \caption{Webpage load time vs. number of articles.}
  \vspace{-0.3cm}
  \label{fig:evalnarticles}
\end{figure}

To better understand the impact that the number of news articles had on performance, we conducted a benchmark test. We used a stripped-down version of an example news website as a basis and created ten different versions of the website, each one with ten more articles than the previous one. The results in Figure~\ref{fig:evalnarticles} show that the overhead by DClaims increases linearly with the number of articles of each page. Surprisingly, Figure \ref{fig:evalwebpage} apparently contradicts this insight: since Sky News' website has more than double the number of news articles than IST's (19 from IST, 42 from Sky News), we expected the former to take longer to load than the latter; however, the opposite occurs. We found that the determining factor in the website's loading time is the Javascript embedded on that page. This is noticeable in NYT's website, where the standard deviation is exceptionally high due to its substantial amount JavaScript code.

\subsection{Performance of IPFS and Ethereum}
\label{sec:perfbackend}

\begin{figure}[t]
  \centering
  \includegraphics[width=0.75\columnwidth]{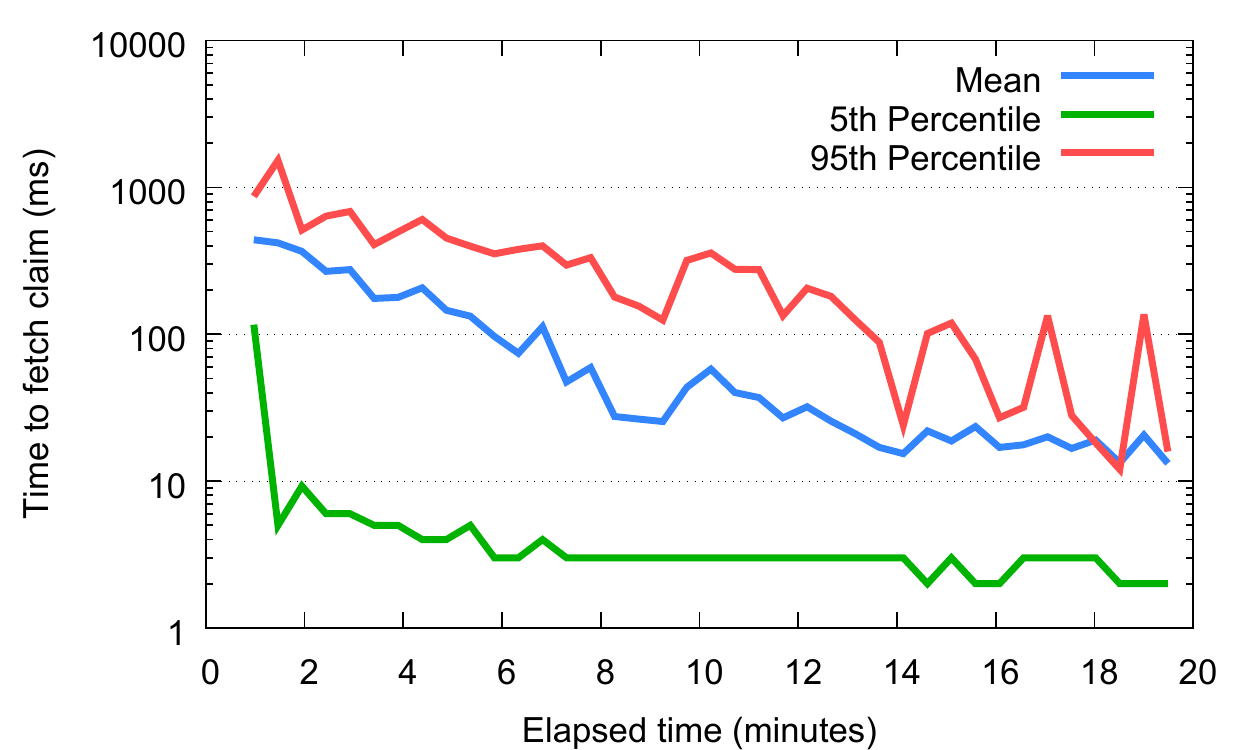}
  \vspace{-10pt}
  \caption{Time to fetch individual claims from IPFS}
  \vspace{-0.3cm}
  \label{fig:ipfsfetch}
  \end{figure}
\begin{figure}[t]
  \centering
  \includegraphics[width=0.75\columnwidth]{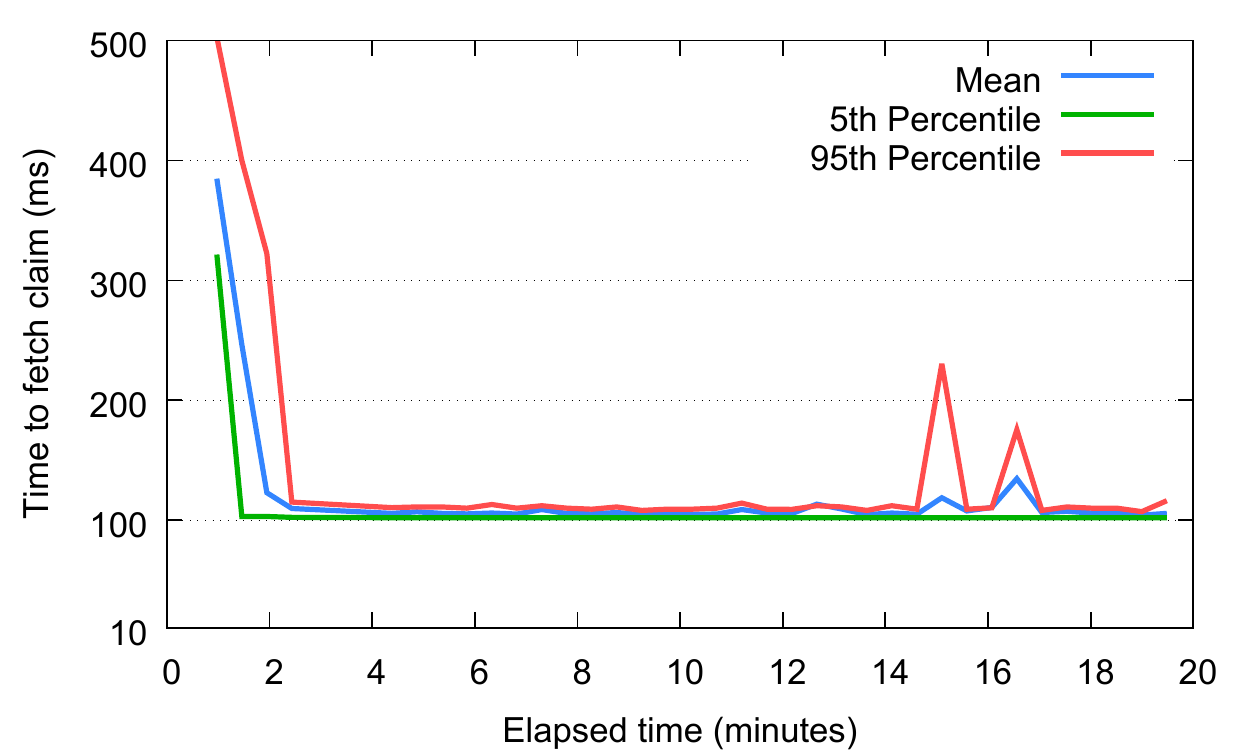}
  \vspace{-10pt}
  \caption{Time to retrieve claim list from Ethereum.}
  \vspace{-0.3cm}
  \label{fig:ethereumfetch}
\end{figure}

IPFS and Ethereum are a crucial part of DClaims since claims are persistently maintained in these systems. For that reason, we aimed to evaluate their performance concerning the time it takes to retrieve claims. Note that the evaluation of the time it takes to issue claims is not as relevant since most of the time is spent on waiting for Ethereum's transaction confirmation and not DClaims' operations. The details of the experimental setup are presented in Section~\ref{eval:ux}. We clarify only that process of retrieving claims from DClaims, for a given article, starts by retrieving the list of claim IPFS links (where retrieving each link corresponds to an Ethereum call) followed by retrieving each claim from IPFS. To evaluate the time claims take to be retrieved, we measured the time it takes, for a given article, to get a list of claim links from Ethereum as well as the time it takes to retrieve individual claims from IPFS.

Figures~\ref{fig:ipfsfetch} and~\ref{fig:ethereumfetch} present our results. Figure~\ref{fig:ipfsfetch} shows that the time to fetch claims from IPFS decreases over time. After elapsed 4 minutes, the claim fetching time falls below 100ms. This trend is explained by the fact that, over time, more claim replicas get cached across IPFS nodes which means that fetching requests can be served from local or nearby IPFS nodes therefore reducing latency. Figure \ref{fig:ethereumfetch} displays the time to retrieve the list of claims from Ethereum. 
In the first minute of the experiment, the values are remarkably higher than in the rest of the elapsed time. We attribute this to Geth (Ethereum daemon) establishing connections to other nodes in preparation for querying them for information. After having established the connections to other nodes, the requests to Ethereum take about 100ms to be completed. The outliers in minutes 15 and 17 were due to some node's Geth losing connection to other nodes. This occurs due to a bug in the Geth client's sync mechanism chosen by us. To reduce AWS costs, we used a faster, experimental, syncing mechanism, which is sometimes unstable and crashed the daemon. The system had to recover from the crash. The same would not occur in a production scenario, as the full sync method would be used. From these results, we find that in their steady states, individual requests to IPFS and Ethereum can be served efficiently with a latency of around 100ms.

\section{Cost Analysis}
\label{eval:facebook}

In this section we analyze the costs of a full-scale deployment of DClaims, which essentially correspond to the costs of sustaining the publishers network. We start, in Section \ref{eval:facebook-data}, by estimating the level of activity that our system would have to endure, using Facebook data to model the potential workload. Next, in Section \ref{eval:cost-publisher}, we calculate the costs of publisher resources based on the considered activity level. Finally, in Section \ref{eval:cost-summary}, we provide an analysis of the cost of the system, offering an example as to how it compares to real-world systems in use today.

\subsection{Analysis of News Pages on Facebook}
\label{eval:facebook-data}

We analyzed Facebook data to learn the level of activity our system would have to support.
In our system, users perform a task similar to the one of commenting on Facebook posts. 
For this reason, we decided to use the rate of interactions (comments, likes, reactions) on four of the most active Facebook's News organization pages to model the activity level that we expect in our system. In other words, we analyzed the rate of interaction that end-users have with these pages, and aim to use these activity levels to estimate the cost and performance of our system on a real setting.

\begin{table}[]
\centering
\begin{small}
\begin{tabular}{@{}lc@{}}
\hline
\multicolumn{2}{c}{\textbf{Facebook Data}}                            \\ \hline
Time duration analyzed (hours)                     & 24      \\
Number of pages analyzed                           & 4       \\
Average number of followers per page (Millions)    & 27      \\
Total number of posts analyzed                     & 200     \\
Total number of interactions                       & 1214549 \\
Average number of posts (per page, per day)        & 50      \\
Average number of interactions (per post, per day) & 6073    \\
Average number of interactions (per page, per day) & 303637  \\
Average length of comment (characters)             & 148     \\ \hline
\end{tabular}
\end{small}
\vspace{0.2cm}
\caption{Activity of news pages on Facebook}
\label{tab:activity-facebook}
\vspace{-0.7cm}
\end{table}

In terms of methodology, we used the Facebook Graph API~\cite{Facebook:wg} to analyze the posts of four of the largest Facebook news pages (CNN, Fox News, The New York Times and BBC News), which averages 27 million users each, during 24 hours (a limitation of the API). For each Facebook post on these pages, we obtained all the comments and a count of the number of likes and reactions. 

From the collected dataset we calculated the values in Table~\ref{tab:activity-facebook}. It shows that, on average, each news Facebook page posts 50 news articles per day, and each post receives around 6073 interactions per day. This means that the activity level that DClaims needs to support, per day, per news organization, is 303637 interactions, which corresponds to the activity level per post, multiplied by the number of posts.

\subsection{Estimation of DClaims Costs}
\label{eval:cost-publisher}

From the values presented in Table \ref{tab:activity-facebook}, we estimate the resources consumed by a publisher in order to withstand the load of a news website. 
We assumed that the volume and rate of web annotations received per news outlet equals those of the news Facebook pages. 
We make the calculations for a single news organization and separate our estimations into the three types of paid resources: storage, computing power and Ethereum transactions.

\mypara{Storage:} To calculate the necessary storage of our system, we started by calculating the storage cost of individual annotations. We used the data format from the W3C Web Annotation standard, and placed a string with 148 characters (average length of Facebook comment) inside each annotation. We calculated that each claim occupied 30KB of storage. From that point, we computed the storage cost per year using AWS pricing for EBS General Purpose SSD\cite{Amazon:ADCIPHn6}. 
This resulted in the value of USD 2203 in storage costs.

\mypara{Computation:} To calculate the computation costs, we started by calculating the number of requests our system would have to process, and then stress-tested a publisher server to evaluate how many servers would be necessary per news outlet. From there we could calculate the annual cost. Our server was able to handle 25 requests per second. We used Table~\ref{tab:activity-facebook}'s \textit{Average number of interactions (per post, per day)} to estimate the number of requests per second we would receive per news article, to then see how many active articles one server would support. We used Table \ref{tab:activity-facebook}'s \textit{Average number of interactions (per post, per day)} to estimate the number of requests per second we would receive per news article, to then see how many active articles one server would support. With each server supporting 1500 requests per minute (25 per second), and the interaction level per article being 4.2 per minute, each server could serve around 357 active articles simultaneously. This number is well above the \textit{Average number of posts (per page, per day)} from Table \ref{tab:activity-facebook}, which means that one server can serve the number of requests expected per news website. The server used was Amazon Web Service's T2.2xlarge, which costs USD 1880 per year\cite{Anonymous:5jQz-EWU}.

\begin{table}[]
\centering
\begin{small}
\begin{tabular}{ll}
\hline
\multicolumn{2}{c}{\textbf{Fixed Values}}      \\ \hline
Ethereum Fiat Value (USD)               & 631  \\
Publisher's Batch Size                  & 100  \\
Transaction Gas Price (Gwei, nanoEther) & 3    \\ \hline
\multicolumn{2}{c}{\textbf{Results}}           \\ \hline
Batch filling time (min)                & 23   \\
Transaction Confirmation Time (min)     & 5    \\
Price per Transaction (USD)             & 0,25 \\ \hline
\end{tabular}
\end{small}
\vspace{0.2cm}
\caption{Ethereum price calculations}
\label{tab:ethereum-prices}
\vspace{-0.9cm}
\end{table}

\mypara{Ethereum transactions:} To calculate the costs on Ethereum transactions, three variables had to be fixed: (1) the value of Ether (the Ethereum blockchain's currency) in fiat currency, USD (value consulted on April 26th, 2018, from~\cite{Coinmarketcap:lIZcYcOd}), (2) the size of the batch the publishers would use, and (3) the Gas Price for the Ethereum transaction confirmation. The selected values and results are shown in Table~\ref{tab:ethereum-prices}. 
%
To choose the gas price for our calculations, we used a popular website for Ethereum consumer metrics, EthGasStation~\cite{EthGasStation:e39VsyQ0} (consulted on April 26th, 2018). We chose the standard value, 3 Gwei, for it is a good compromise between waiting time and price. 30 minutes was too long to wait, and 5 Gwei was too expensive. 
%
we know from Table~\ref{tab:activity-facebook} that the number of interactions per article is 4.2 per minute, which means a batch of size 100 would take 23 minutes to fill and would reduce the transaction cost per claim by a factor of 100 (when compared to one transaction per claim). 23 minutes per batch with a 100 fold reduction in cost is a good compromise, so the selected batch size for this test was 100.
Lastly, from the gas price (3 Gwei) and batch size (100) values, we can calculate the yearly cost in Ethereum transactions, which is of USD 277069.

\subsection{Cost Analysis Findings} 
\label{eval:cost-summary}

Table~\ref{tab:costs-final} presents the main findings of our study. Running the DClaims system for one big news outlet, such as CNN, Fox News, BBC News or The New York Times, would approximately cost USD 281152 per year. This value was calculated for only one of these large news outlets, so the value presented does not represent the real world cost. However, even if we assume that around the world there are 30 news outlets the size of the ones analyzed, DClaims' costs are still significantly smaller than the ones for real-world systems with a donation based financial model, such as Wikipedia.

\begin{table}[]
\centering
\begin{small}
\begin{tabular}{ll}
\hline
\multicolumn{2}{c}{\textbf{Final Costs}}    \\ \hline
Storage                            & 2203   \\
Computation                        & 1880   \\
Ethereum                           & 277069 \\
Total cost for 1 year              & 281152 \\ \hline
\multicolumn{2}{c}{\textbf{Useful Metrics}} \\ \hline
Cost per 1000 Claims (USD)         & 2,54   \\
Cost per User for 2,7M users (USD) & 1,041  \\ \hline
\end{tabular}
\end{small}
\vspace{0.2cm}
\caption{Final cost analysis per news outlet}
\label{tab:costs-final}
\vspace{-0.9cm}
\end{table}

In perspective, these values are very reasonable. The Facebook news pages analyzed have, on average, 27 million users. Even if we assume that DClaims only attracts 1\% of those users the cost per user, per news outlet, would be less than USD 1 per year. 
DClaims targets users who need to circumvent censorship. Many people pay monthly fees for security services such as VPNs, ranging from USD 5 to 10 per month, which equals USD 60 to USD 120 per year. Therefore, if a donation based financial model such as Wikipedia does not succeed, there is reason to believe a subscription-based service, would. Thus, we infer there is both a market and a viable financial model for third-parties willing to host DClaim publishers.

\section{Related Work}
\label{sec:rel}

We present the related work on areas concerning the building blocks of DClaims and in censorship-resistant content dissemination.

\noindent {\bf Web annotation services:}
Genius Web Annotator~\cite{genius} is a web annotation system that allows for the annotation of any webpage, provided that the user, or the webpage, supports the Genius plugin. Hypothes.is~\cite{hypo} is similar to Genius, but implements a standard data model of web annotations (from W3C), which increases interoperability with other services. The main limitation of these services is that all the data is stored on their servers, meaning they offer no guarantees of persistence or availability. Moreover, having full control over data means they can engage in censorship. Both of these issues are addressed in DClaims.

\noindent {\bf IPFS-based systems:}
IPFS is an open-source project, with thousands of contributors and in use in many decentralized projects, such as market places~\cite{OpenBazaar:wh} and identity systems~\cite{Anonymous:91thCMME}, among others. IPFS has also empowered censorship resistance efforts in the past, one example being copying Wikipedia to IPFS, making it available in Turkey, when the government blocked the website~\cite{Anonymous:jpQ4Givt}. To our knowledge, IPFS has not yet been used to store and transport web annotations. Using IPFS for providing censorship resistant access to web annotations is a good use of the system. However, there still needs to be a registry to keep track of all this data, which is where Ethereum smart-contracts become useful.

\noindent{\bf Blockchain technology:} The blockchain technology was introduced by Bitcoin~\cite{Anonymous:JOJGrvgg,Bonneau:2015ema,Gervais:2014bs}. Since then, there has been a lot of research on blockchain and cryptocurrencies, mostly focused on new consensus protocols~\cite{Anonymous:-Ouaq-Dl,Milutinovic:2016gs,Bentov:2014ws,Dziembowski:2015gs,Eyal:2016vn}, vulnerability assessment~\cite{Bonneau:2015ema,Gervais:2016dd,Atzei:2017ju}, privacy~\cite{Anonymous:zwRGR7mQ,Kosba:2016iq,Biryukov:2015kg} and some in social applications ~\cite{Tomescu:2017iz,Ali:2016vq}. Around the Ethereum blockchain has also borne an ecosystem of applications that have used for many purposes~\cite{Dapps:cEzXcv-U}. The Ethereum applications that most relate to our work are the ones which use smart-contracts as data registries, this is the case with uPort~\cite{uPort:hYIek7hR}, a decentralized identity platform and Ethereum Name Service~\cite{ens:MMvDWWWa}, a DNS-like system built on top of Ethereum. The novelty in DClaims' use of Ethereum is in the way scalability is dealt with in the application layer, using the publisher network. Other scalability solutions for Ethereum, such as sharding~\cite{EthereumFoundation:blLahP4N} and side-chains~\cite{Poon:2018uu} and state-channels~\cite{Network:2018ta}, operate at lower levels.

\noindent{\bf Censorship-resistant content dissemination:}
Many decentralized systems provide access to documents, implementing replication and redundancy algorithms to circumvent censorship attempts by content deletion~\cite{Clarke:2000fy,Waldman:2001ema,Waldman:2000wf,Serjantov:2002fj}. 
In Freenet~\cite{Clarke:2000fy}, files are split into chunks and spread out through several servers on the network. The network is made of volunteered servers, each server offering some amount of storage space. Tangler~\cite{Waldman:2001ema} operates in a similar way. The key difference between these systems is that Tangler makes it harder for servers to delete content deliberately. This is because each file that is added to the network contains blocks of other files already hosted. The result is that is a server deletes a certain file, he will be affecting many other files, making it easy for the network to detect the attack. Publius~\cite{Waldman:2000wf} also spreads files through a network of servers, but since its main goal is privacy, files are encrypted. In contrast to DClaims, these systems offer no disincentives to misbehavior. Marty et al.~\cite{Marty:2018uc} attempt to solve this issue by introducing a cryptographic proof of censorship, but it requires all the content stored to be encrypted, which is undesirable in the context of DClaims, as everyone should have access to the information. DClaims offers a different way to deal with bad actors.

\section{Conclusion}
\label{chap:conclusion}

This paper presents DClaims, a decentralized web annotations platform which is resistant to censorship. DClaims stores data in a distributed network and keeps a registry of metadata on the Ethereum blockchain, which is a tamper-proof, permanent record of information. 
To address the limitations of blockchain technology, DClaims uses a small network of dedicated nodes called publishers. 
We built a reference implementation of the system on the form of a browser extension, which allows for the web annotation of news websites, allowing users to classify news articles, and view the classifications made by others. Our evaluation shows DClaims can support the same level of activity of Facebook's news organizations pages. 

\bibliographystyle{ACM-Reference-Format}
\bibliography{main}

\end{document}